\documentclass{article}
\usepackage[utf8]{inputenc}
\usepackage[margin=2cm]{geometry}
\usepackage[table]{xcolor}

\title{A computational approach to aid clinicians in selecting anti-viral drugs for COVID-19 trials}

\author{Aanchal Mongia$^{1}$, Sanjay Kr. Saha$^{2}$, Emilie Chouzenoux$^{3*}$ \& Angshul Majumdar$^{4*}$\\ \\
        $^{1}$Dept. of CSE, IIIT - Delhi, India, 110020\\
	   	$^{2}$Department of Community Medicine, IPGMER Kolkata\\
       	$^{3}$CVN, Inria Saclay, Univ. Paris Saclay, 91190 Gif-sur-Yvette, France\\
       	$^{4}$Dept. of ECE, IIIT - Delhi, India, 110020\\
        $^{*}$Corresponding authors contact/Email:\\ angshul@iiitd.ac.in, emilie.chouzenoux@centralesupelec.fr\\}
        
\date{3 July, 2020}

\usepackage[numbers]{natbib}
\usepackage{graphicx}

\usepackage{url}
\usepackage{comment}
\usepackage{mathtools}
\usepackage{tikz}

\usepackage{makecell}
\usepackage{soul}

\begin{document}

\maketitle




\begin{abstract}
\textbf{Motivation:}
COVID-19 has fast-paced drug re-positioning for its treatment. This work builds computational models for the same. The aim is to assist clinicians with a tool for selecting prospective antiviral treatments. Since the virus is known to mutate fast, the tool is likely to help clinicians in selecting the right set of antivirals for the mutated isolate. 

\textbf{Results:} 
The main contribution of this work is a manually curated database publicly shared, comprising of existing associations between vi\-ruses and their corresponding antivirals. The database gathers similarity information using the chemical structure of drugs and the genomic structure of viruses. Along with this database, we make available a set of state-of-the-art computational drug re-positioning tools based on matrix completion. The tools are first analysed on a standard set of experimental protocols for drug target interactions. The best performing ones are applied for the task of re-positioning antivirals for COVID-19. These tools select six drugs out of which four are currently under various stages of trial, namely Remdesivir (as a cure), Ribavarin (in combination with others for cure),  Umifenovir (as a prophylactic and cure) and Sofosbuvir (as a cure). Another unanimous prediction is Tenofovir alafenamide, which is a novel tenofovir prodrug developed in order to improve renal safety when compared to the counterpart tenofovir disoproxil. Both are under trail, the former as a cure and the latter as a prophylactic. These results establish that the computational methods are in sync with the state-of-practice. We also demonstrate how the selected drugs change as the SARS-Cov-2 mutates over time, suggesting the importance of such a tool in drug prediction.

\textbf{Availability:} The dataset along with the solution is made available publicly at \url{https://github.com/aanchalMongia/DVA} and the prediction tool named DVApred (Drug-virus association prediction server) with a user-friendly interface is available as a webserver at \url{http://dva.salsa.iiitd.edu.in}.\\ 


\textbf{Contact:} angshul@iiitd.ac.in, emilie.chouzenoux@centralesupelec.fr

\end{abstract}

\section{Introduction}
There has been an exponential rise in the total active cases and deaths due to COVID-19 (COrona VIrus Disease-2019) since the first case in Wuhan, China in December, 2019 \cite{worldometer}. The disease results in severe acute respiratory syndrome coronavirus 2 (SARS-CoV-2), which is known to be highly transmittable and has spread across more than 100 countries. This pandemic has wreaked havoc on people's social life, the global economy, and most importantly the health of the human race. The death numbers are frightening, confirming about 467K deaths worldwide till mid-June, 2020 \cite{worldometer}.

As medical professionals are striving to save lives, research scientists specialized in drug development, are racing against time to develop a vaccine against SARS-CoV-2 \cite{harrison2020coronavirus}. The investigation involved for developing a vaccine (or even a new drug) is time consuming, requiring several phases of extensive trials. Experts believe that it is highly unlikely that a vaccine will be ready before a year or more. In such circumstances the best bet may be to re-position existing drugs for treating COVID-19. This is a well known approach where existing drugs (which have already been approved for release in the market) are investigated for new disease/s \cite{ashburn2004drug}. Drug re-positioning is usually cost effective and fast (compared to developing a new drug / vaccine) since its effects are well studied. One classic example for drug re-positioning is Chlorocyclizine , which was initially developed as an anti-allergic but later found to act against the hepatitis C virus \cite{he2015repurposing}. Another example is Imatinib mesylate (sold under the trade name Gleevec), it was originally used as a treatment for leukemia but later was found to be effective against genes associated with gastrointestinal-stromal tumors \cite{mclean2005imatinib, frantz2005drug}.  

Given the relatively large drug-virus association space, manual investigation in wet-labs is not a scalable strategy. Putting all the anti-virals in trials for treating corona is not very feasible either; especially because time is of essence. In such a scenario, computational approaches can help; they can be used to prune down the search space for the drugs to be investigated \cite{survey1}. {Practically, such approaches could also assist the clinicians to come up with treatments for rapidly mutating viruses by pruning the anti-viral drug space (see Discussion section). Specifically, a computational approach which takes into account the genomic structure of the latest viral isolate or its similarity with the previously occuring strains of viruses would be helpful in deciding the treatment}. With this objective, we have manually curated a comprehensive database called DVA (Drug Virus Association), having the approved (anti-viral) drug-virus associations in the literature along with the similarity measures associated with drugs (chemical structure similarity) and viruses (genome sequence similarity). To the best of our knowledge there is no existing database for drug virus association. 

The DVA database we propose in this work lies the foundation for further computational studies on this topic. There can be various methodologies to predict drug virus association. The prediction problem can be approached via feature-based classification models, neighborhood models, matrix completion models, network diffusion models etc. A recent empirical study on well established drug-target interaction databases exhibit the best prediction performance by matrix completion models \cite{survey1}. In computer science, matrix completion is used routinely in recommendation systems
. The general problem of drug-disease association can actually be thought of as a recommendation system, where drugs are being recommended for treating a disease. 
Given the success of matrix completion techniques in drug target interaction, we deploy state-of-the-art matrix completion techniques on our curated DVA database. We perform a thorough comparative analysis of those for predicting assessed drug-disease associations. Then, we apply the methodology for pruning the search space of potential candidates for COVID-19 trial drugs. Finally, we show how the tool helps in selecting drugs as the virus mutates.

Our objective is to make our solution user friendly for clinicians and scientists. In pursuit of this goal, we have made the solution (dataset and algorithms) available as a webserver. The webserver can be used in two ways. First, given the genome of the virus, the webserver can predict (and rank) the probable antivirals. Second, given a drug and a virus, it can output a normalized score depicting how effective the drug will be against the virus. 

\section{Results}
We assess the performance of different matrix completion techniques in this section. The techniques have been described in the Methods section. Six matrix completion methods were used, which can be categorized into three families provided below.
\begin{itemize}
    \item Basic frameworks (MF: Matrix factorization \cite{mongia2019mcimpute} and MC: Matrix completion or Nuclear norm minimization \cite{mongia2019mcimpute})
    \item Deep frameworks (DMF: Deep matrix factorization) \cite{mongia2020deep},
    \item Graph regularized frameworks (GRMF: graph regularized matrix factorization \cite{ezzat2017drug}, GRMC: graph regularized matrix completion \cite{mongia2020drug}, GRBMC: graph regularized binary matrix completion \cite{mongia2020computational})
\end{itemize}  


Matrix factorization (MF) is the traditional matrix completion method which factorizes the data matrix into two latent factor matrices (tall and short) and the algorithm recovers these factor matrices to recover the original matrix. Since this problem is non-convex, it may not converge to a global minimum of the cost function. Nuclear-norm minimization based matrix completion (MC) was proposed as a (mathematically) better alternative; it directly recovers the matrix by penalising its nuclear norm (convex surrogate of rank). Deep matrix factorization (DMF) generalises MF to more than two factors. None of the techniques mentioned so far can take advantage of genomic structure of the viruses or chemical structure of the drugs. The said pieces of information can be incorporated into the graph regularized matrix completion techniques (GRMF, GRMC, GRBMC). These techniques have been explained in detail in the Methods section.

\subsection{Overview: DVA prediction}
The typical anti-viral drug discovery process involves genomic and biophysical understanding of the virus. It aims to target the enzymes or peptides involved in the viral replication cycle and takes years for successful clinical validation. Other approaches involve screening all the broad-spectrum anti-viral drugs or chemical libraries comprising large numbers of existing compounds/databases (having information on transcriptional signatures in different cell lines) to be further evaluated by standard anti-viral assays \cite{zumla2016coronaviruses}. In view to assist acceleration of this process (by pruning down the search space), we create and share a publicly available DVA database, along with a number of matrix completion techniques (mentioned above) for drug-virus association prediction. 

\begin{figure}[h!]
\centering
\includegraphics[width=.98\linewidth]{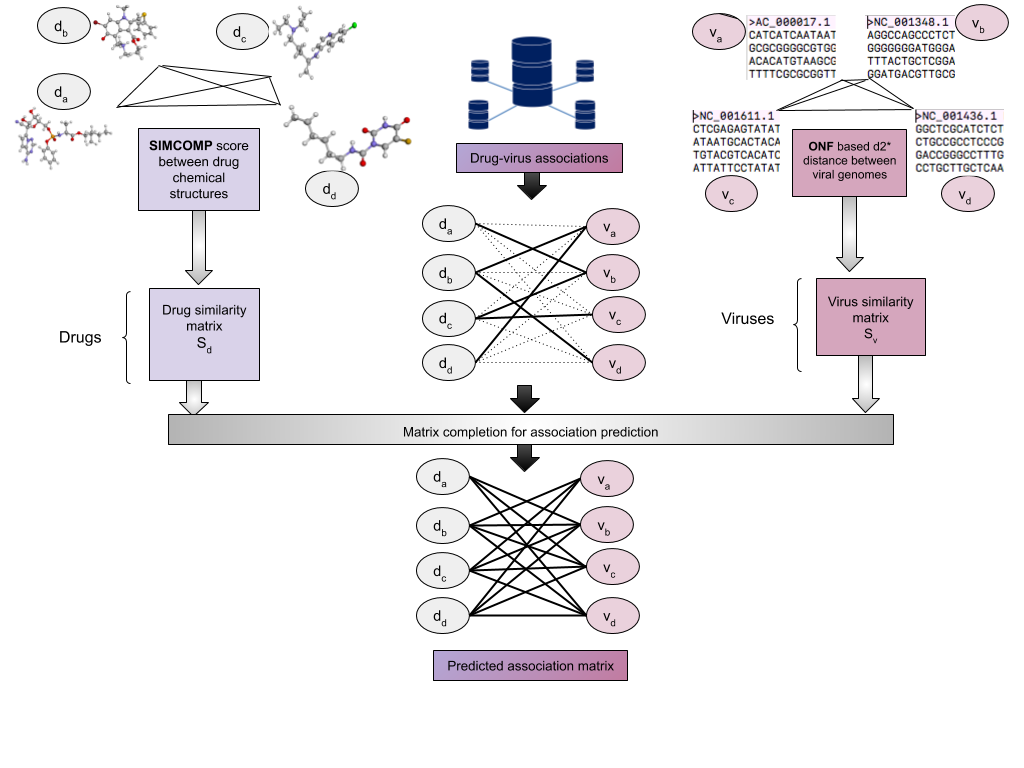}
\caption{Schematic diagram depicting the DVA framework}
\label{fig:pipeline}
\end{figure}

The originality of the proposed work lies in the formalization of the drug-virus association prediction as a matrix completion problem, without the need for any anti-viral assays. Such a computational approach requires the chemical structure of the drugs and, in case of graph-regularized matrix completion techniques, the genome of the viruses, or existing associations otherwise. Figure \ref{fig:pipeline} depicts the schematic flow of the proposed work involving data curation and implementation overview.

\subsection{Empirical Evaluation}
\label{sec:2.2}

In this sub-section, we carry out extensive experimental protocol to illustrate and compare the ability of the different methods to retrieve the drug-disease associations available in our curated dataset. The protocol dictates three variants of 10-fold cross-validation setting (CV). In the first setting CV1 (cross validation 1), 10 \% of the associations selected at random are left out as testing set. This allows to assess each algorithm's ability to predict associations between existing drugs and viruses. To evaluate an algorithm for its ability to predict association for novel drugs and viruses i.e. those which have no association information, we use two other (more stringent) CV settings. In CV2 and CV3, 10 \% of the complete virus and drug entities selected at random are left out as test set respectively.

The standard metrics for evaluation are the AUC (Area under the Receiver Operating Characteristic curve) and AUPR (Area under the precision-recall curve). AUC is more common in machine learning literature, it assumes that the classes are evenly balanced. Problems in drug-disease association have highly imbalanced classes, in such a scenario the AUPR is a more appropriate metric for evaluation~\cite{burez2009handling,ezzat2017drug}.

Table \ref{t:10fcv} shows how each of the 6 tested algorithms performs in retrieving the associations. A clear observation from the experiments is that the graph regularized-based matrix completion algorithms that incorporate the similarity information associated with the drugs and viruses, perform fairly well giving an AUC greater or equal than 0.83 in CV1. The best performing algorithm (GRBMC) exhibits an AUC and AUPR of 0.88 and 0.54 respectively. Predicting the associations for novel drugs and viruses also have a reasonable performance with the best AUC/AUPR of 0.81/0.44 and 0.73/0.31 by GRBMC and GRMF. It can be noted that the standard matrix completion methods, which do not take into account the metadata, fail to learn from the association data giving a near-random performance as far as the prediction on novel viruses is concerned, depicting how very important the similarity information is.

\begin{table}[h!]
\centering
\begin{tabular}{l|l|llllll}
\hline \hline
&Metric   & MC    &MF    &DMF         &GRMF     &GRMC& GRBMC\\
\hline
\hline
CV1&AUC       &0.5959	&0.6753	&0.6974	&0.8652		&0.8279	&0.8834\\
&AUPR       &0.3238 	&0.2656	&0.2615	&0.4812		&0.4445	&0.5220\\
\hline 
\\
\hline
CV2&AUC       &0.4909	&0.5033	&0.5704	&0.7346		&0.6705	&0.6632\\
&AUPR       &0.1106	&0.0504	&0.0855	&0.3112		&0.2951	&0.2746\\ 
\hline
\\
\hline
CV3&AUC       &0.5438	&0.5215	&0.4529	&0.7806		&0.7507	&0.8181\\
&AUPR       &0.0538	&0.0637	&0.0824	&0.4265		&0.4333	&0.4383\\
\hline
\hline
\end{tabular}

\caption{\textbf{Results for association prediction for all techniques under the 3 cross validation settings}.}
\label{t:10fcv}
\end{table}

\subsection{Association prediction for new drugs}
\label{sec:2.3}
DVA database and its associated computational tools can also be used on new drugs without any previously known virus association information. For evaluating this ability, we identified in our database all the drugs which are known to interact with only one virus (drugs associated with a single virus only) and hide that association to the methods. This allows us to assess the performance of the algorithms in predicting viruses associated with the new drugs in the database.

We hide the only virus corresponding to each of the 76 drugs (with only a single virus associated with it) and run matrix completion to predict candidate viruses for these drugs. The drugs for which the test virus associated with it is the top-ranked virus predicted by the algorithm would have the maximum precision value (MPV) of 1. The number and percentage of drugs with a maximum precision value of 1 are reported in Table \ref{t:newDrugPred}.

Nearly 34 \% (26/76) of single association drugs with a maximum precision of 1 were predicted using GRMF. Other graph regularized frameworks show comparable performance in terms of predicting drugs with MPV of 1.

\begin{table}[h!]
\centering

\begin{tabular}{l|lllllll}
\hline \hline
   & MC    &MF    &DMF         &GRMF        &GRMC & GRBMC\\
\hline
\hline
\# drugs with MPV=1  &2    &4    &4      &26       &22 &8\\
\hline
\% drugs with MPV=1  &2.6316 &5.2632 &5.2632 &34.2105 &28.9474 &10.5263\\
\hline
\hline
\end{tabular}
\caption{\textbf{Number and percentage of drugs predicted with MPV=1 by the matrix completion methods}.}
\label{t:newDrugPred}
\end{table}

\subsection{SARS-CoV-2 prediction}
\label{sec:2.4}
In this experiment, we add the SARS-CoV-2 sample in our database by providing its ONF based d2* similarity \textbf{\cite{ahlgren2017alignment}} in the virus similarity matrix. 

We then apply the matrix completion algorithms to predict the associations and rank prediction scores corresponding to SARS-CoV-2 to predict the top 10 recommended drugs.

\begin{table}[h!]
\centering

\begin{tabular}{l|l}
\hline \hline
\textbf{Technique} & \textbf{SARS-Cov-2} \\
\hline
\hline

			
GRMF &   \cellcolor{yellow}\textcolor{blue}{Remdesivir}	\\	\cline{2-2}
&	\cellcolor{yellow}\textcolor{blue}{Ribavirin}	\\	\cline{2-2}
&	\cellcolor{yellow}\textcolor{blue}{Sofosbuvir}	\\	\cline{2-2}
&	\cellcolor{yellow}\textcolor{red}{Umifenovir}	\\	\cline{2-2}
&	\textcolor{blue}{Taribavirin}	\\	\cline{2-2}
&	\textcolor{blue}{Tenofovir alafenamide}	\\	\cline{2-2}
&	\cellcolor{yellow}Ibuprofen	\\\cline{2-2}
&	Pleconaril	\\\cline{2-2}
&	Geldanamycin	\\\cline{2-2}
&	\textcolor{blue}{Vidarabine}	\\\cline{2-2}
			
\Xhline{2\arrayrulewidth}			
			
			
GRMC&	\cellcolor{yellow}\textcolor{blue}{Remdesivir}	\\	\cline{2-2}
&	\cellcolor{yellow}\textcolor{blue}{Ribavirin}	\\	\cline{2-2}
&	\cellcolor{yellow}\textcolor{blue}{Sofosbuvir}	\\	\cline{2-2}
&	\textcolor{blue}{Taribavirin}	\\\cline{2-2}
&	\textcolor{blue}{Tenofovir alafenamide}	\\	\cline{2-2}
&	\textcolor{blue}{Vidarabine}	\\\cline{2-2}
&	Telaprevir	\\	\cline{2-2}
&	Boceprevir	\\	\cline{2-2}
&	Simeprevir	\\	\cline{2-2}
&	Palivizumab	\\	\cline{2-2}
			
\Xhline{2\arrayrulewidth}			
			
			
GRBMC&	\cellcolor{yellow}\textcolor{blue}{Remdesivir}	\\	\cline{2-2}
&	\cellcolor{yellow}\textcolor{blue}{Ribavirin}	\\\cline{2-2}
&	\cellcolor{yellow}\textcolor{blue}{Sofosbuvir}	\\\cline{2-2}
&	\cellcolor{yellow}\textcolor{red}{Umifenovir}	\\	\cline{2-2}
&	\textcolor{blue}{Taribavirin}	\\\cline{2-2}
&	\textcolor{blue}{Vidarabine}	\\\cline{2-2}
&	Brivudine	\\	\cline{2-2}
&	\textcolor{blue}{Tenofovir alafenamide}	\\	\cline{2-2}
&	Paritaprevir	\\	\cline{2-2}
&	Peginterferon alfacon-1	\\	\cline{2-2}

\Xhline{2\arrayrulewidth}

\end{tabular}
\caption{\textbf{Top-10 drugs predicted for SARS-Cov-2 by the DVA computational methods}.}
\label{t:top5}
\end{table}

As can be seen from the results of section \ref{sec:2.2} (Table \ref{t:10fcv}), MC, MF and DMF often yield considerably worse results than their graph regularized counterparts (GRMF, GRMC and GRBMC). Such poor performance of non-graph regularized versions of matrix completion methods could be explained as they do not incorporate any knowledge about the genomic structure of the viruses and the chemical structure of the drugs. Since the three graph-based methods perform reasonably well in the prediction task, we consider these techniques for the drug prediction on the novel coronavirus. The top-10 drugs they predicted have been reported in Table \ref{t:top5} (ranked by their predicted scores). Drugs highlighted with blue text are unanimously predicted drugs by the three considered matrix completion techniques and those in red text are predicted by two methods. We also highlighted with yellow cells the drugs which are under trial/investigation as a potential cure/prophylactic against COVID-19.


It can be seen that the three techniques have consistently and unanimously selected six drugs, namely Remdesivir, Ribavarin, Sofosbuvir, Taribavirin, Tenofovir alafenamide and Vidarabine. Umifenovir has been recommended by two (GRMF and GRBMC) out of three techniques. Amongst these recommendations, Remdesivir \cite{beigel2020remdesivir}, Ribavarin \cite{hung2020triple,zeng2020comparative}, Sofosbuvir \cite{rodrigo2020sofosbuvir} and Umifenovir \cite{wang2020clinical} are under clinical trials. Taribavirin is similar to Ribavirin but it is not approved by the FDA. Tenofovir alafemanide (an antiretroviral for HIV-1) is on undergoing trial \cite{duan2020advance}. GRMF has additionally selected Ibuprofen which is expected to be investigated in UK~\cite{ibuprofenTrialUK,martins2020no}. The fact that three techniques unanimously select the aforementioned drugs make us confident about these recommendation results.

\subsection{Predictions evolution with mutating novel coronavirus}
\label{sec:2.5}
In the previous sub-section, we have established that the results from our models are mostly in sync with clinical practice. In this sub-section, we will demonstrate how our proposed approach can be of help to clinicians. 

All the results generated so far have been generated using the reference sequence of the SARS-Cov-2 strain (collected in December, 2019 in Wuhan). The novel coronavirus is rapidly mutating \cite{chatterjee2020overview}. In such a scenario, it is necessary to select drugs that are effective against the mutated strain. While mutating, the virus isolates may develop resistance to previous drugs used for its treatment. Our model may be of help to clinicians in this respect. Before proposing a treatment regime (trial, for e.g.) for COVID-19 treatment, the practitioner may use our approach to check the drugs selected for the particular isolate of novel coronavirus. 
In Table \ref{t:mutatingIsolates}, we have experimented with three isolates of the novel coronavirus (collected over an interval of 2 months), in addition to the reference sequence (collected in December 2019). Those three isolates have been collected in February (from USA), April (from Australia) and June (from India).

One can note from the Table \ref{t:mutatingIsolates} that the selected drugs change with mutations. Baloxavir marboxil was not selected even once for the reference sequence from December 2019, but has been selected by two methods for the February isolate. A recent pre-print \cite{lou2020clinical} reports the results of this antiviral on COVID-19 patients. The drug Ibuprofen, was selected by one of the methods for the December reference sequence, it was not selected for the February isolate, then it was selected by two methods for the April isolate and selected by all three for the June isolate. It may be worthy to note that lipid Ibuprofen is being considered in a trial in UK from starting June, 2020 \cite{ibuprofenTrialUK}. Similarly, Pleconaril has been selected for by all three methods for the most recent (June) isolate, it was selected by only one of the techniques for the reference sequence (December) and was not selected for the February or June sequences. Pleconaril, although developed for treating enterovirus and rhinovirus, is not FDA approved. Rilpivirine and Etravirine are two antiretrovirals developed for treating HIV positive subjects. Both of them have been predicted by all three methods in the latest isolate, but not in the previous isolates or in the reference sequence. To the best of our knowledge, this antiretroviral is not under study for COVID-19 trials. Note that Vidarabine, which was getting predicted for the reference sequence (albeit wrongly) has not been predicted from the later ones. 
Based on this discussion, we can see that how the mutations in genomic structure results in different predictions of drugs. Since the novel coronavirus is mutating, it may be judicious to account for the structure of the latest isolate while deciding the treatments to be put in trial. In such a case, our model may be of help to clinicians.

\begin{table}[h!]
\centering

\begin{tabular}{l|lll}
\hline \hline
\textbf{Technique} & \textbf{SARS-Cov-2: february} & \textbf{SARS-Cov-2: april}  & \textbf{SARS-Cov-2: june}\\
\hline
\hline

							
GRMF &	Remdesivir	&	Remdesivir	&	Remdesivir	\\	\cline{2-4}
&	Ribavirin	&	Sofosbuvir	&	Umifenovir	\\	\cline{2-4}
&	Umifenovir	&	Umifenovir	&	Pleconaril	\\	\cline{2-4}
&	Taribavirin	&	Ribavirin	&	Ibuprofen	\\	\cline{2-4}
&	Sofosbuvir	&	Tenofovir alafenamide	&	Sofosbuvir	\\	\cline{2-4}
&	Baloxavir marboxil	&	Ibuprofen	&	Rilpivirine	\\	\cline{2-4}
&	Geldanamycin	&	Pleconaril	&	Etravirine	\\	\cline{2-4}
&	Tenofovir alafenamide	&	Hydroxychloroquine	&	Tenofovir alafenamide	\\	\cline{2-4}
&	Tecovirimat	&	Valomaciclovir	&	Rimantadine	\\	\cline{2-4}
&	Peramivir	&	Dexamethasone	&	Ribavirin	\\	\cline{2-4}
							
\Xhline{2\arrayrulewidth}							
							
							
GRMC&	Remdesivir	&	Remdesivir	&	Umifenovir	\\	\cline{2-4}
&	Umifenovir	&	Sofosbuvir	&	Remdesivir	\\	\cline{2-4}
&	Ribavirin	&	Tenofovir alafenamide	&	Ibuprofen	\\	\cline{2-4}
&	Taribavirin	&	Boceprevir	&	Pleconaril	\\	\cline{2-4}
&	Sofosbuvir	&	Telaprevir	&	Sofosbuvir	\\	\cline{2-4}
&	Vidarabine	&	Palivizumab	&	Chloroquine	\\	\cline{2-4}
&	Tenofovir alafenamide	&	Simeprevir	&	Etravirine	\\	\cline{2-4}
&	Nelfinavir	&	Ribavirin	&	Rilpivirine	\\	\cline{2-4}
&	Amprenavir	&	Umifenovir	&	Tenofovir alafenamide	\\	\cline{2-4}
&	Boceprevir	&	Ibuprofen	&	Nelfinavir	\\	\cline{2-4}
							
\Xhline{2\arrayrulewidth}							
							
							
GRBMC &	Remdesivir	&	Remdesivir	&	Umifenovir	\\	\cline{2-4}
&	Ribavirin	&	Umifenovir	&	Remdesivir	\\	\cline{2-4}
&	Umifenovir	&	Sofosbuvir	&	Pleconaril	\\	\cline{2-4}
&	Taribavirin	&	Ribavirin	&	Ibuprofen	\\	\cline{2-4}
&	Sofosbuvir	&	Taribavirin	&	Sofosbuvir	\\	\cline{2-4}
&	Paritaprevir	&	Paritaprevir	&	Rilpivirine	\\	\cline{2-4}
&	Tenofovir alafenamide	&	Brivudine	&	Etravirine	\\	\cline{2-4}
&	Atazanavir	&	Vidarabine	&	Ribavirin	\\	\cline{2-4}
&	Baloxavir marboxil	&	Daclatasvir	&	Tenofovir alafenamide	\\	\cline{2-4}
&	Favipiravir	&	Beclabuvir	&	Trifluridine	\\	\cline{2-4}
							
\Xhline{2\arrayrulewidth}

\end{tabular}
\caption{\textbf{Top-10 drugs predicted for three isolates of SARS-Cov-2 (collected at an interval of 2 months) by the DVA computational methods}.}
\label{t:mutatingIsolates}

\end{table}

\subsection{Execution time}
We recorded the time taken by each of the matrix completion algorithms for a single run (Table \ref{t:time}), on a single core machine with a clock speed of 2.8 GHz, 64 GB RAM (Intel(R) Xeon(R) CPU E5-1603 v3 processor).
All the methods have relatively low computational requirements. Matrix factorization methods are faster than the nuclear norm minimization based techniques, with a difference of few seconds. Such difference may not be practically significant, given the nature of our problem as an improved anti-viral prediction in pandemic is much more crucial than the running time in the order of seconds.


\begin{table}[h!]
\centering
\begin{tabular}{l|lllllll}
\hline \hline
   & MC    &MF    &DMF       &GRMF    &GRMC &GRBMC\\
\hline
\hline
Time (sec)  &0.0859 & 0.0149 &0.0529  & 0.0457 &10.55 & 5.22\\
\hline
\hline
\end{tabular}
\caption{\textbf{Running time of the DVA computational methods}.}
\label{t:time}
\end{table}

\section{Discussion}
We have collected a comprehensive dataset comprising of all the anti-viral drugs which act against viruses known to infect humans, along with the similarity information associated with the drugs and the viruses (see Methods section). On this database, we deploy state-of-the-art drug target interaction techniques based on matrix completion. 

\subsection{General Discussion}
The drug-virus associations and the similarity information are assembled as three matrices: drug-virus association matrix ($Y$), drug similarity matrix ($S_d$) and virus similarity matrix ($S_v$). Several matrix completion methods have then been implemented and compared. The matrix completion methods which are not designed to incorporate the similarity information take association matrix as input (assuming it to be a partially filled matrix from which the full low-rank association matrix would be recovered) along with the masking operator which stores information on the position of train and test indices. On the other hand, the graph regularized frameworks utilize similarity information and give an improved prediction performance in the cross-validation evaluation (Best AUC=0.88, AUPR=0.52). The graph regularized matrix completion methods are not only capable of predicting associations between existing drugs and viruses but can also take into account novel viruses and drugs for which no association information is known (as can be seen in the latter two cross-validation settings). The similarity information for such novel drugs/viruses ($S_v$ and $S_v$) can be added to the metadata using the chemical structure and sequence information of the drug and virus respectively. The validity of the proposed pipeline is illustrated by the fact that 4 out of the 6 drugs unanimously predicted in top-10 prediction by the graph regularized methods are already under trial for treating SARS-Cov-2. 

\subsection{Drug recommendations for the novel coronavirus}
From the prediction provided by graph regularized methods, we observe a consensus over the recommendation of six drugs, namely Remdesivir, Ribavarin, Sofosbuvir, Taribavirin, Tenofovir alafenamide and Vidarabine (7th common drug being Umifenovir recommended by 2 models). 
Note that Umifenovir and Remdesivir are investigational as per FDA, Umifenovir is approved in Russia and China and Remdesivir has obtained approval for emergency use by FDA. Researchers working on Ribavarin trials argue that since it is an established drug with ready availability and established supply chains, it is worth to investigate its potency against COVID-19.
In a study published in April, 2020, Remdesivir was found to shorten the recovery time (median recovery time of 11 days, 31 \% faster recovery time than patients who received placebo) in adults with COVID-19 infection in clinical trials involving 1063 hospitalized patients in United states \cite{beigel2020remdesivir}. It has received Emergency Use Authorization (EUA) by the US Food and Drug Administration (FDA) for patients hospitalized with severe disease \cite{us2020fact}.
Umifenovir (arbidol) is also being investigated as a potential prophylactic agent for the prevention of COVID-19 \cite{wang2020clinical}. 
Ribavirin \cite{hung2020triple,zeng2020comparative}, in combination with other antiviral drugs has recently been studied for the effectiveness and safety of different antiviral regimens (combination therapies) for the treatment of COVID-19. 
Sofosbuvir is used specifically for hepatitis C infection. Currently it is under trial for treating COVID-19 patients. This is because, superposition of the hepatitis C virus polymerase bound to sofosbuvir, with the SARS-CoV polymerase shows that the residues that bind to the drug are present in the latter \cite{rodrigo2020sofosbuvir}. 

 The clinical trial of Tenofovir disoproxil as a prophylactic, is based on recent albeit sparse literature that shows that RNA synthesis nucleos(t)ide analogue inhibitors, acting as viral RNA chain terminators, like Tenofovir disoproxil, abacavir or lamivudine, amongst others, could have an effect against SARS-CoV-2 \cite{tenofovirDisoproxilTrials}. Our algorithm selects Tenofovir alafenamide, which is less harmful to the kidneys than Tenofovir disoproxil. Tenofovir alafenamide is known to have large antiviral efficacy at ten times lesser dose than Tenofovir disoproxil. It is also under investigation as a combination therapy (emtricitabine/tenofovir-alafenamide and lopinavir/ritonavir) to treat COVID-19 patients \cite{duan2020advance}.

\subsection{Symptomatic treatment options from unanimously predicted drugs}

In this sub-section, our objective is to establish that the drugs selected by our algorithms for SARS-Cov-2 are clinically sensible predictions, in the sense that they are known to be effective against a significant number of the COVID-19 symptoms. The Table \ref{t:top5} shows that out of the top ten selections, six are common across all the three techniques and another (Umifenovir) have been predicted by two out of three. Although Vidarabine has been selected by all three models, it is an antiviral effective against DNA viruses. Since the novel coronavirus is an RNA virus this antiviral is not supposed to work. We have discussed the rest of the six antivirals in \textbf{Supplementary section 1}. The descriptions have been primarily taken from \url{drugbank.ca}. 

\begin{figure}[h!]
\centering
\includegraphics[width=.95\linewidth]{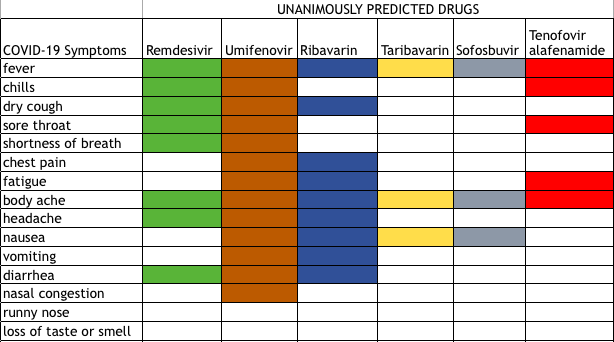}
\caption{List of COVID-19 symptoms treated by drugs unanimously predicted by the three graph-regularized matrix completion methods.}
\label{fig:symptoms_drugs}
\end{figure}

Symptoms of different virus infections have been discussed in \textbf{Supplemetary section 2}. From a symptomatic aspect, Umifenovir, Remdesivir and Ribavarin covers most of the symptoms for COVID-19. All of them are now under clinical trials as discussed in the previous subsection.

\subsection{Effect of viral mutations on treatments}

Furthermore, significance of such computational models for predicting anti-viral therapeutics would correlate with the rate at which a virus mutates. RNA viruses (like HIV, flu virus) are known to mutate at a much faster rate than the DNA viruses \cite{sanjuan2010viral}, helping them to evade the human immune system and develop drug resistance. SARS-CoV-2 is no exception and has been mutating over the past few months \cite{pachetti2020emerging}. Clinically keeping up with the evolving viruses and drug resistance could be a major challenge for development of an anti-viral treatment \cite{hussain2017drug}. Hence, artificial intelligence, and in particular the presented matrix completion techniques, could help the clinicians to prune down the drug space for viral strains which have been mutating rapidly and avoid unnecessary testing on drugs for the new viral strain/s.

\subsection{Conclusion}
Computational techniques have the inherent advantage of learning from the data (which can be huge) and scale to a large number of drugs and viruses and hence be of immense importance to the clinicians by narrowing down the search space for the clinical trials to be carried out. 

We would like to emphasize that the proposed DVA database and methods are not particular to the novel coronavirus. Such computational approaches have the general capability to help for identification of drugs which might be effective against a broad spectrum of viruses \cite{huggins1989prospects}, or the viruses which can be targeted by multiple drugs (since many drugs could target specific elements of viral replication) \cite{schaefer2012anti}. We believe that the proposed work will pave the way for more scientific ideas for anti-viral drug re-positioning and assist clinicians in the process.

\section{Methods}
\subsection{Drug-virus associations database}
The proposed DVA dataset aims at being exhaustive. It compiles various existing sources, housing together all the anti-viral drugs proved clinically to be effective against viruses infecting humans. We believe such resource would be highly useful for analysing and proposing anti-virals not only for the novel coronaviruses but other viruses too. Along with that, it may also be used to computationally identify viruses that a newly discovered drug may target. The associated metadata (information about the drugs and viruses) may also help clinicians in manual analysis and having a deeper insight.

All the associations corresponding to anti-viral drugs clinically shown to act against human host viruses have been assembled from standard DrugBank database \cite{wishart2006drugbank} (\url{https://www.drugbank.ca/categories/DBCAT000066}). To ensure that the database is fully comprehensive, other literature works \cite{razonable2011antiviral,de2016approved,sugaya2010long,chopra2014effectiveness,gallegos2016chikungunya,das2014heat,shiryaev2017repurposing,winther2003potential,jin2020structural} and resources such as ViPR \cite{pickett2012vipr} were also scanned for any additional drug-viral associations. ViPR  or NIAID Virus Pathogen Database and Analysis Resource (\url{http://www.viprbrc.org/}) is a repository of data and analysis tools for virology research \cite{pickett2012vipr} capturing various types of information derived from comparative genomics analysis and visualization tools. It has antiviral drug information (for 21 viral species) derived imported from DrugBank (\url{https://www.drugbank.ca/}) \cite{wishart2006drugbank}.

The drug-virus indications have been stored (see \textbf{Supplementary data section}) and processed in a matrix form of size $m \times n$ ($m$ being no of drugs in the database and $n$ being the number of viruses) to be used as input for any of the 6 matrix completion algorithms we made available in our repository. 

The DrugBank Identifier (DrugBank ID) of the anti-viral drugs involved is considered as the unique key for the drugs, obtained from DrugBank vocabulary (\url{https://www.drugbank.ca/releases/latest#open-data}). Along with the viral association information, we also store the target pathway and mechanism of action of each drug for quick reference in any further investigation. Apart from this, each drug is mapped to its corresponding KEGG Identifier (KEGG ID) from the KEGG Compound/KEGG Drug database (\url{https://www.genome.jp/kegg/drug/}, \url{https://www.genome.jp/kegg/compound/}) of the KEGG (Kyoto encyclopedia of genes and genomes) \cite{kanehisa2006genomics}. The KEGG IDs were taken from the linking file provided at \url{https://www.drugbank.ca/releases/latest#external-links} \cite{wishart2006drugbank} or manually added in the case of drugs missing in the linking file. 

Each virus is identified by an acronym assigned to it (in case of no acronym, full virus name is used). The viral family, genome type, transmission route and incubation period is also available in the virus metadata file along with the accession number of the complete genomic sequence of the viruses fetched from NCBI (National Center for Biotechnology Information) Viral genome browser \url{https://www.ncbi.nlm.nih.gov/genomes/GenomesGroup.cgi} \cite{NCBI}).

\subsection{Similarity computation}
To integrate the similarity information to the drug-virus associations, we have computed similarities between the drugs based on their chemical structures and between the viruses using their complete genomic sequences.
\begin{itemize}
    \item DRUG SIMILARITY: All the DrugBank IDs were mapped to KEGG IDs of the corresponding drug/compound in the KEGG database \cite{kanehisa2006genomics}. The chemical structure similarity was measured between the drugs by computing the SIMCOMP score \cite{hattori2010simcomp} based on the maximum common substructures between the chemical structure of the compounds using the KEGG API page at GenomeNet (\url{https://www.genome.jp/tools/gn_tools_api.html}). The drugs for which the SIMCOMP score was less than the set cutoff (0.001) and the drugs with no KEGG IDs available were assigned a similarity score of 1 to themselves and 0 to other drugs in the dataset.

    \item VIRUS SIMILARITY: The $d2^*$ distance based on ONF (Oligonucleotide frequency) measure between the DNA sequences was shown to be the best amongst various other ONF metrics with several $k$-mers length in host prediction accuracy at the genus level \cite{ahlgren2017alignment}. Hence, we compute d2* dissimilarity/distance (at $k$=6) between the viral genome sequences obtained from NCBI \cite{NCBI}. The reference sequences of viruses were saved in FASTA format to be used by the distance computation software (\url{https://github.com/jessieren/VirHostMatcher}) proposed by \cite{ahlgren2017alignment}. The $d2^*$ distance was subtracted from 1 to obtain the similarity measure.
    
    For the viruses with segmented structure (Influenza A virus, Influenza B virus, Influenza C virus, Lassa mammarenavirus), the coding sequence in the nucleotide sequence of each genomic segment (taken in decreasing order of length was taken) was combined to form the complete viral sequence.

\end{itemize}

\subsection{Proposed method: Matrix completion}
In this subsection, we describe each of the matrix completion algorithms used (\textbf{\url{www.github.com/AanchalMongia/DVA}}), along with their mathematical formulations and resolution strategies. 

Let $X_{m \times n}$ be the complete drug-virus association matrix (with $m$ drugs and $n$ viruses) with binary entries (1 denoting that the drug is known to act against the virus and 0 denoting no association). Here $X$ is the matrix to be recovered from its sampled (partially known) entries in $Y$. Let $M$ denote the masking operator (elementwise multiplied to $X$) having 1's at positions where associations are known and 0 otherwise. Then, the matrix completion problem can be formulated as searching for $X$ satisfying:
\begin{equation}\label{eq:1}
Y = M(X),
\end{equation}
under specific constraints. In particular, it is typically assumed that similar drugs act in a similar manner, hence $X$ to be recovered (from $Y$ and $M$) is of low-rank.

\subsubsection{Matrix factorization (MF)}
The most straightforward technique of solving low-rank matrix completion is matrix factorization, where the data matrix $X_{m \times n}$ is decomposed into two latent factor matrices $U_{m \times k}$ and $V_{k \times n}$, where $k$ denotes the number of latent (hidden) factors deciding if a drug is associated with a virus or not. $X$ is recovered by solving for $U$ and $V$ in the following minimization problem:
\begin{equation} \label{eq:3}
\mathop {\min }\limits_{U,V} ||Y - M(UV) ||_F^2.
\end{equation}
The above problem is solved in an alternating manner, by first decoupling the mask using a majorization-minimization technique \cite{major-min,Chouzenoux2013} and then using alternating least squares method \cite{hastie2015matrix} to obtain $U$ and $V$. The complete algorithm is described in \cite{mongia2019mcimpute}.

\subsubsection{Deep matrix factorization (DMF)}
An extension of matrix factorization has been proposed motivated by the success of deep dictionary learning \cite{tariyal2016deep}, where the data matrix $X$ is decomposed into multiple factor matrices (analog to multiple layers) to capture more complex hidden features in the data. The formulation of the minimization problem in the case of 2-layer matrix factorization is given below:
\begin{equation} \label{eq:4}
\mathop {\min }\limits_{U_1,U_2,V} ||Y - M  (U_1U_2V)||_F^2 \text{ such that } U_1\geq 0, U_2\geq0.
\end{equation}
The above problem is solve alternatively. The minimization with respect to variables $U_1$ and $V$, is done in a similar way to that of matrix factorization, while the update on $U_2$ can be obtained as shown in \cite{mongia2019deepmc}.

\subsubsection{Graph regularized matrix factorization (GRMF)}
Another variant of Matrix factorization has been proposed to incorporate metadata associated with the row and column entities (drug and virus similarities in this case) \cite{ezzat2017drug}. Here, the drug and virus entities form the nodes of two separate graphs and the similarity between them is assumed to be the weights between the nodes. Regularization is imposed by adding graph Laplacian penalties to the cost function of matrix factorization as shown below:
\begin{equation}
 \label{eq:5}
\mathop {\min }\limits_{U,V} ||Y - M(UV)||_F^2 + {\mu _1}\rm{tr}(U{^\top}  {L_d} U) + \\{\mu _2} \rm{tr}(V  {L_v} {V^\top}), \hfill 
\end{equation}
where $\mu _1 > 0$ and  $\mu _2 > 0$ are coefficients penalizing the graph regularization Laplacian terms and $\rm{tr}$ denotes the trace of the matrix. $L_d = D_d - S_d$ and $L_v = D_v - S_v$ are the graph Laplacians \cite{chung1996spectral} for $S_d$ (row/drug similarity matrix) and $S_v$ (column/virus similarity matrix), respectively, and $D_d^{ii}=\Sigma_j S_d^{ij}$ and $D_v^{ii} = \Sigma_j S_v^{ij}$ are the associated degree matrices. A resolution technique for the above formulation has been shown in \cite{ezzat2017drug}.

\subsubsection{Matrix completion (MC)}
Matrix factorization based approach leads to a non-convex minimization problem and hence rarely benefits from global convergence guarantees. To limit the space of minimizers, it may be useful to impose a low-rank constraint on the solution $X$. Since rank minimization is still an NP-hard problem, it was proposed to relax the above constraint to its closest convex surrogate  by making use of the nuclear norm penalty \cite{near-optimal,exact}. The formulation for the resulting nuclear norm minimization problem (referred to as matrix completion by the authors) is:
\begin{equation} \label{eq:7}
\mathop {\min }\limits_X ||X|{|_*}  \text{ such that } \textit{Y=M(X) }.
\end{equation}
 The above problem can be solved alternatively, by invoking majorization-minimization arguments \cite{major-min} to deal with the mask operator $M$ and by applying thresholding operations on the singular values to process the nuclear norm term \cite{mongia2019mcimpute}.

\subsubsection{Graph regularized matrix completion (GRMC)}
Just like matrix factorization, nuclear norm minimization based matrix completion can also be graph regularized by incorporating graph Laplacian penalties to take  metadata/similarity information into account. The formulation for the minimization problem is given by:
\begin{equation}
\label{eq:8}
\mathop {\min }\limits_{X} ||Y - M(X)||_F^2 +\lambda ||X||_* + {\mu _1}\rm{tr}(X^\top  {L_{d}} X) + {\mu _2}\rm{tr}(X  {L_{v}} {X^\top}). 
\end{equation}
The above formulation can either be solved using ADMM (alternating direction method of multipliers) \cite{boyd2011alternating,Komodakis} as was done in \cite{mongia2020drug} (referred as GRMC here) or by explicitly taking care of the constraint that the recovered values should be in the range $[0,1]$. If the latter range constraint is taken into account, we obtain then a new variant called graph regularized binary matrix completion. The minimization with respect to $X$ can be solved by making use of the PPXA (parallel proximal algorithm)~\cite{pustelnik2011parallel}. Such approach allows to decouple the constraints by  introducing proxy variables and then solving each subproblem in a parallel fashion as shown in \cite{mongia2020computational} (referred as GRBMC here).

\subsection{Setting of hyperparameters}
The stepsize, regularization parameters and latent factor dimensions, for the above techniques have been tuned using cross-validation on training set (after hiding 10 \% of the data) in each of the three cross-validation settings (see Section \ref{sec:2.2}). The parameters obtained after extensive cross-validation on the setting CV2 (randomly hiding the virus entities) have been further used in predicting drugs for SARS-Cov-2 and the corresponding isolates (see Sections \ref{sec:2.4} and \ref{sec:2.5}). Similarly, the parameters selected for the setting CV3 (randomly hiding drug entities) have been used to evaluate the performance of the approaches in Section \ref{sec:2.3}.



\clearpage

\begin{center}
\textbf{\large SUPPLEMENTARY:\\ A computational approach to aid clinicians in selecting anti-viral drugs for COVID-19 trials }
\end{center}

\setcounter{section}{0}
\renewcommand{\thesection}{\arabic{section}}

\section{Description of drugs predicted by computational models}
\begin{itemize}

\item \textbf{Remdesivir} (FDA emergency use, investigational) – Remdesivir was originally investigated as a treatment for Ebola virus, but has potential to treat a variety of RNA viruses. Its activity against the coronavirus (CoV) family of viruses, such as SARS-CoV and MERS-CoV, was shown in 2017, and it is also being investigated as a potential treatment for COVID-19.

\item  \textbf{Ribavirin} (approved) – Broad-spectrum activity against several RNA and DNA viruses. It is primarily indicated for use in treating hepatitis C and viral hemorrhagic fevers. It is reported that ribavirin might be only effective in early stages of viral hemorrhagic fevers including Lasser fever, Crimean-Congo hemorrhagic fever, Venezuelan hemorrhagic fever, and Hantavirus infection. Currently Ribavirin is being used in combination with interferon beta-1b, lopinavir–ritonavir in a trial for treating COVID-19.  

\item  \textbf{Umifenovir} (Investigational) – Umifenovir is used for the treatment and prophylaxis of influenza and other respiratory infections. Umifenovir's ability to exert antiviral effects through multiple pathways has resulted in considerable investigation into its use for a variety of enveloped and non-enveloped RNA and DNA viruses, including Flavivirus / hemorrhagic fever, Zika virus, Lassa virus, Ebola virus, chikungunya virus, Hantaan virus, and coxsackie virus A and B. Umifenovir is currently being investigated as a potential treatment and prophylactic agent for COVID-19.



\item  \textbf{Sofosbuvir}– Sofosbuvir is a direct acting antiviral medication used as part of combination therapy to treat chronic Hepatitis C. Currently it is undergoing clinical trial for treating COVID-19. 

\item  \textbf{Taribavirin }(similar to Ribavirin) – The prodrug taribavirin is under development for the treatment of patients with chronic hepatitis C. Taribavirin is metabolized by the liver and converted into its active metabolite, ribavirin. This pathway reduces exposure to red blood cells (RBCs) and increases exposure to the liver, the site of HCV replication.

\item  \textbf{Tenofovir alafenamide }– Tenofovir alafenamide is a novel tenofovir prodrug developed in order to improve renal safety when compared to the counterpart tenofovir disoproxil. Tenofovir alafenamide is indicated to treat chronic hepatitis B, treat HIV-1 and prevent HIV-1 infections. Currently tenofovir disoproxil is undergoing trial as a prophylactic against COVID-19 on healthcare workers. The study is supposed to be completed in July. 

\end{itemize}

\section{Symptoms of different virus infections}

\begin{itemize}

\item \textbf{Hemorrhagic fever symptoms (Mayoclininc, Clevelandclinic)} – fever, fatigue, dizziness, body ache, headache, rashes, bleeding from the eyes, ears or mouth, difficulty breathing, internal bleeding, organ failure. 
\\\textbf{Drugs:} Ribavarin, Umifenovir
\\ \\
\textbf{SARS-CoV symptoms (CDC, WHO)} – fever, dry cough, sore throat, shortness of breath, headache, body ache, loss of appetite, malaise, night sweats, chills, confusion, rash and diarrhea. 
\\\textbf{Drugs:} Remdesivir
\\ \\
\textbf{MERS-CoV symptoms (CDC, WHO)} – fever, chills, headache, body ache, malaise, shortness of breath, diarrhea, dry cough, sore throat, body ache and hypoxia. 
\\\textbf{Drugs:} Remdesivir
\\ \\
\textbf{Influenza virus (CDC)} – fever, body ache, chills, sweats, headache, dry cough, fatigue, nasal congestion and sore throat. 
\\\textbf{Drugs:} Umifenovir, Ibuprofen
\\ \\
\textbf{Zika virus (CDC)} - mild fever, rash, body ache, headache, red eyes and malaise. 
\\\textbf{Drugs:} Umifenovir
\\ \\
\textbf{Lassa virus (WHO)} – fever, fatigue, malaise, headache, sore throat, body ache, chest pain, nausea, vomiting, diarrhoea, cough and abdominal pain. 
\\\textbf{Drugs:} Umifenovir, Ribavarin
\\ \\
\textbf{Ebola virus (CDC, WebMD)} – fever, body ache, headache, abdominal pain, rash, loss of appetite, fatigue, diarrhea, vomiting, unexplained hemorrhaging, bleeding or bruising. 
\\\textbf{Drugs:} Remdesivir, Umifenovir
\\ \\
\textbf{Chikungunya virus (CDC)} – fever, body ache, headache, joint swelling and rash. 
\\\textbf{Drugs:} Umifenovir
\\ \\
\textbf{Hantavirus (CDC)} – fatigue, fever, body ache, headache, dizziness, chill, nausea, vomiting, diarrhea, abdominal pain, blurred vision, inflamed or red eyes, shortness of breath, rash, low blood pressure. Drugs: Umifenovir, Ribavarin
coxsackie virus A (WebMD) – sore throat, blisters in mouth, and small tender lesions on the palms of their hands and bottom of their feet, inflammation of the spinal cord and brain. 
\\\textbf{Drugs:} Umifenovir
\\ \\
\textbf{Coxsackie virus B (WebMD)}  – fever, spasms of the abdominal and chest muscles, inflammation of the spinal cord and brain. 
\\\textbf{Drugs:} Umifenovir
\\ \\
\textbf{Non-polio Enterovirus (CDC)} – fever, runny nose, sneezing, cough, skin rash, mouth blisters, body muscle aches. 
\\\textbf{Drugs:} Pleconaril
\\ \\
\textbf{Rhinovirus  (Mayoclinic)} – runny nose, sore throat, cough, congestion, mild body aches,  mild headache, sneezing, low fever and malaise. 
\\\textbf{Drugs:} Ribavarin, Sofosbuvir, Taribavirin
\\ \\
\textbf{Hepatitis B (DNA virus)} (hepb.org) – fever, fatigue, body ache, loss of appetite, nausea, vomiting,
stomach pain, pale or light colored stools, dark urine, jaundice, bloated stomach.
\\\textbf{Drugs:} Tenofovir alafenamide
\\ \\
\textbf{HIV-1 (acute phase)} – fever, chills, rash, night sweats, body ache, sore throat, fatigue, swollen lymph
nodes, mouth ulcers.
\\\textbf{Drugs:} Tenofovir alafenamide

\end{itemize}

\section{Readme for Webserver}

    
The web server (\url{http://dva.salsa.iiitd.edu.in}) has two tabs in the menu. The `Home' page provides reference to our work and the main functionalities of the web server. The `About' page consists of a brief description of the dataset and the algorithms used.
\\ \\
\textbf{Home} – The `Home' page briefly discusses what this web server is about. It also includes reference to our pre-print. In the sub-section `Functionalities', there are two parts. The first one pertains to the prediction of drugs, given the genomic structure of the virus. The second takes in two inputs, a drug and a virus, and it returns a normalized score depicting the overall efficacy of the drug against the virus, as predicted by our computational model. 

\begin{figure}[h!]
\centering
\includegraphics[width=.98\linewidth]{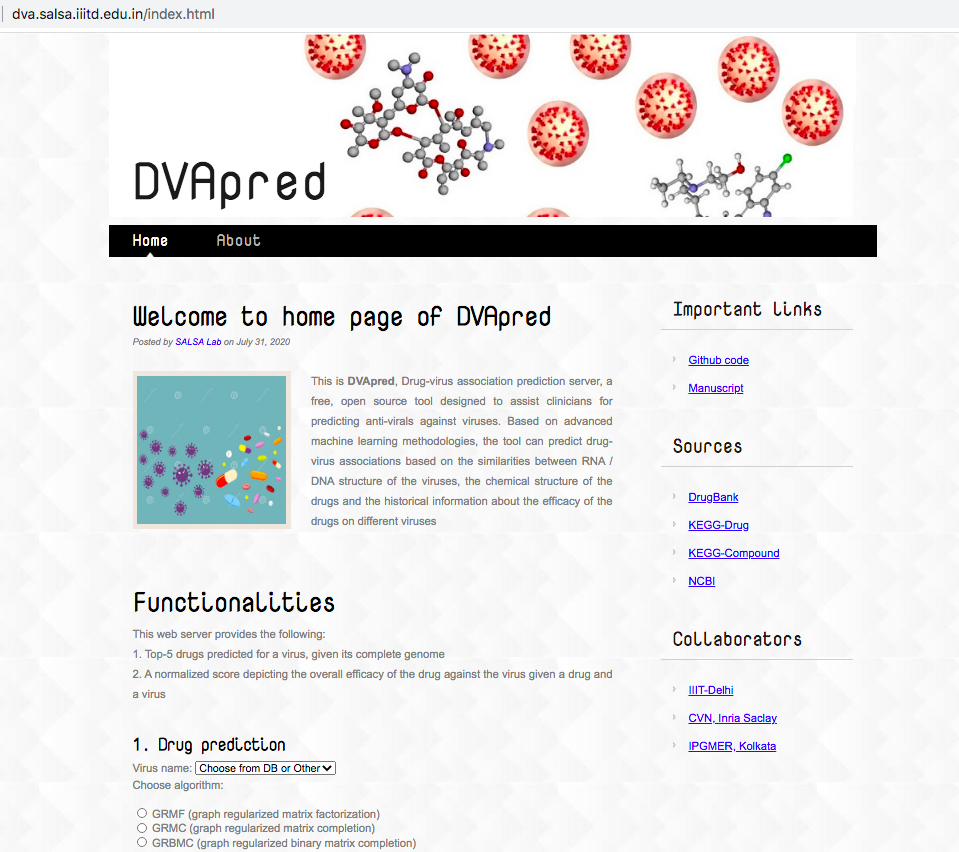}
\caption{Home page of the web-server}
\label{h1}
\end{figure}

\begin{figure}[h!]
\centering
\includegraphics[width=.98\linewidth]{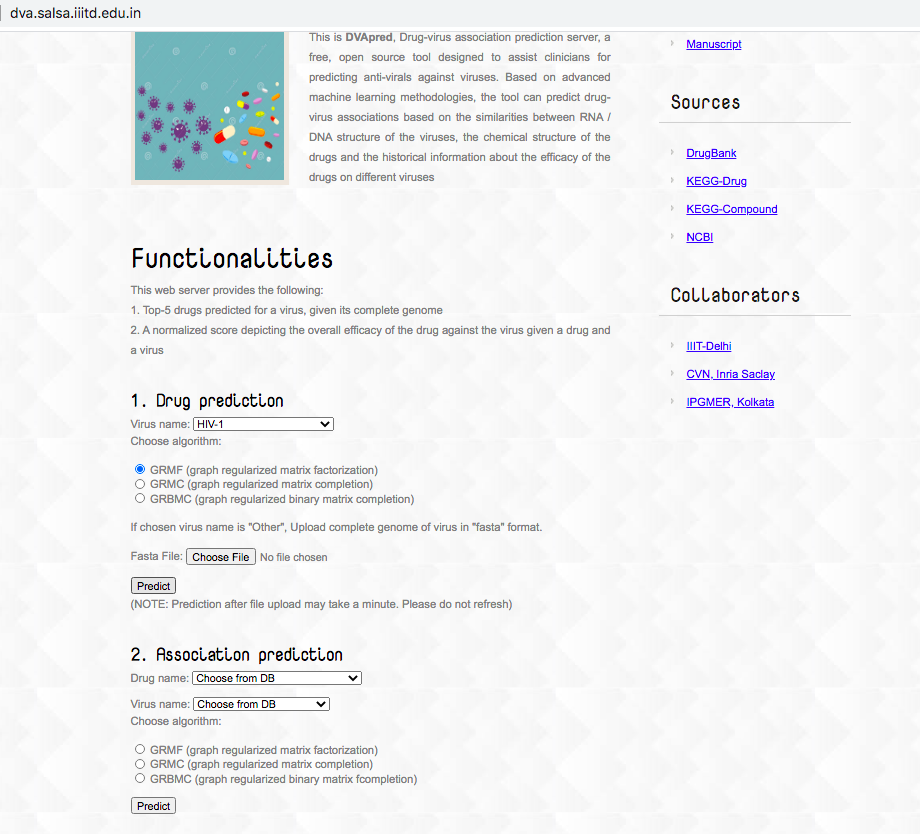}
\caption{Sample query 1- To predict top-5 drugs for a virus already existing in the database }
\label{h2}
\end{figure}

\begin{figure}[h!]
\centering
\includegraphics[width=.98\linewidth]{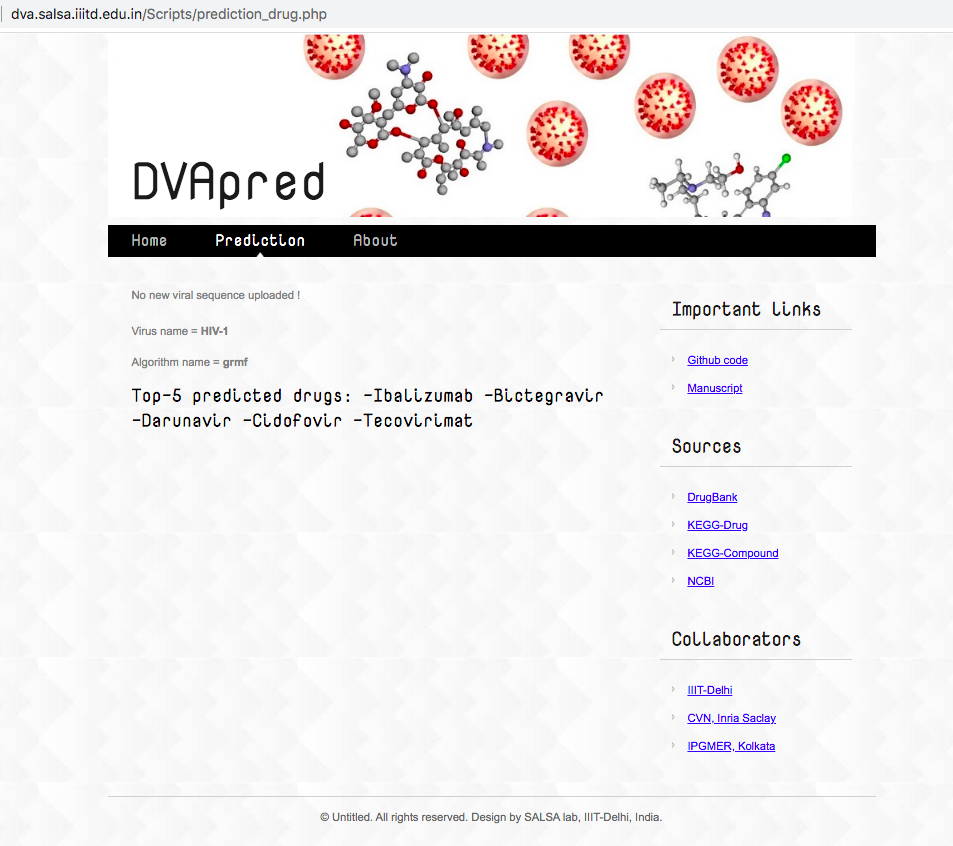}
\caption{Sample query 1- Prediction results}
\label{h3}
\end{figure}

\begin{figure}[h!]
\centering
\includegraphics[width=.98\linewidth]{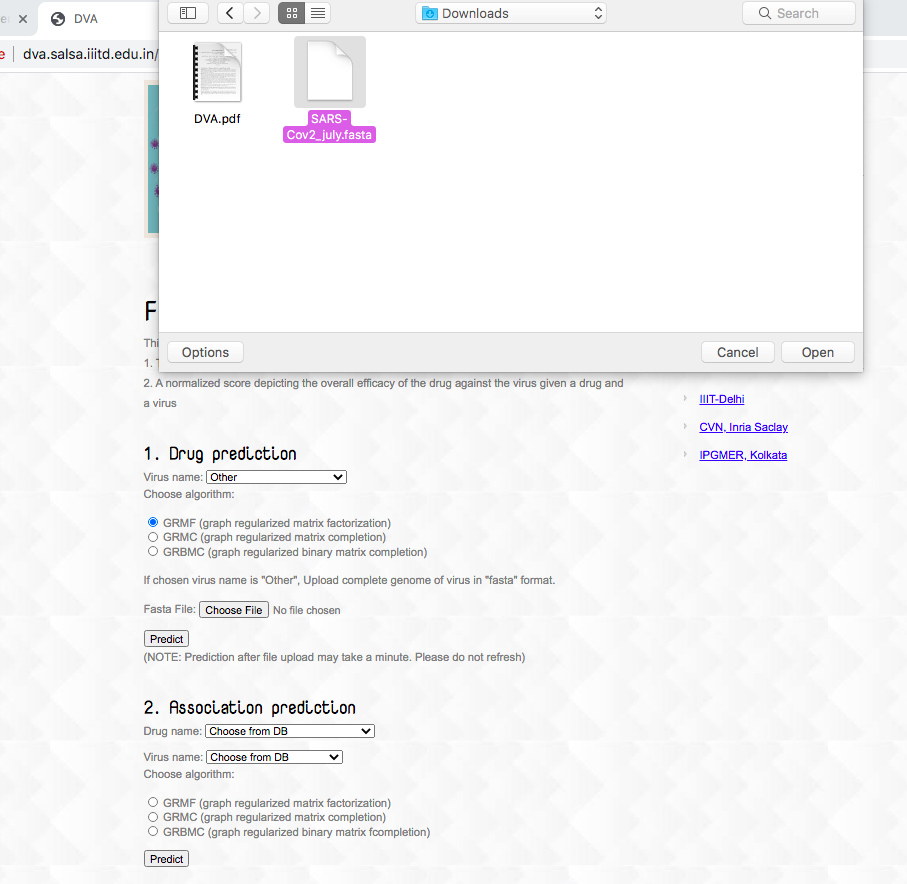}
\caption{Sample query 2- To predict top-5 drugs for a novel virus by uploading its genomic sequence}
\label{h4}
\end{figure}

\begin{figure}[h!]
\centering
\includegraphics[width=.98\linewidth]{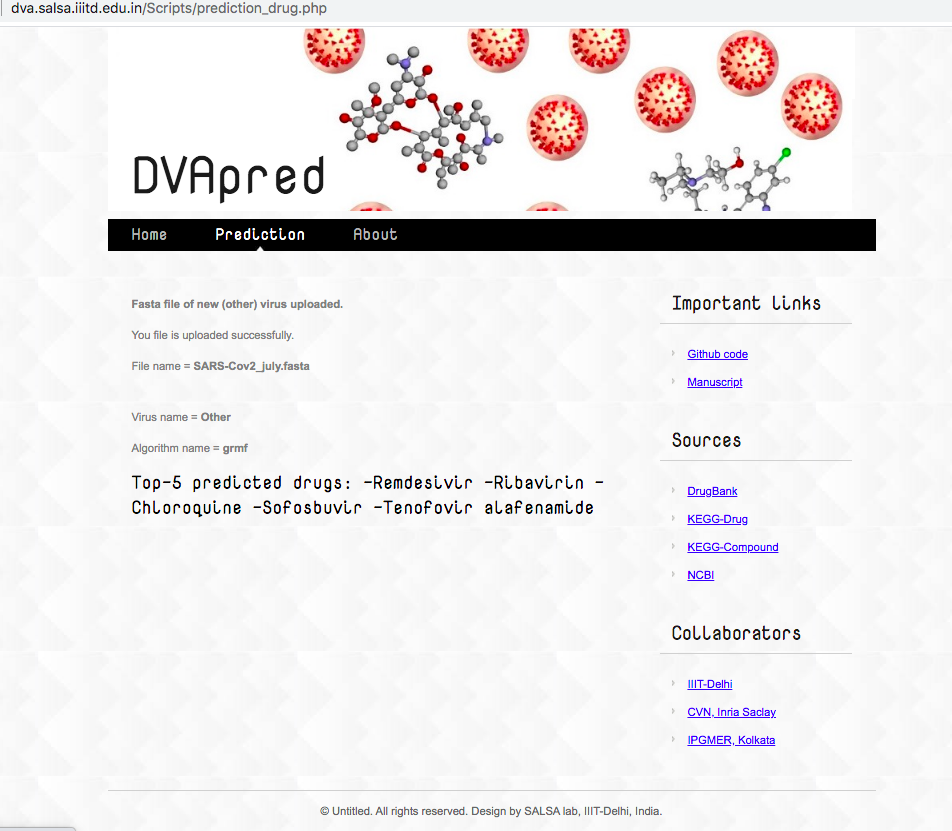}
\caption{Sample query 2- Prediction results}
\label{h5}
\end{figure}

\begin{figure}[h!]
\centering
\includegraphics[width=.98\linewidth]{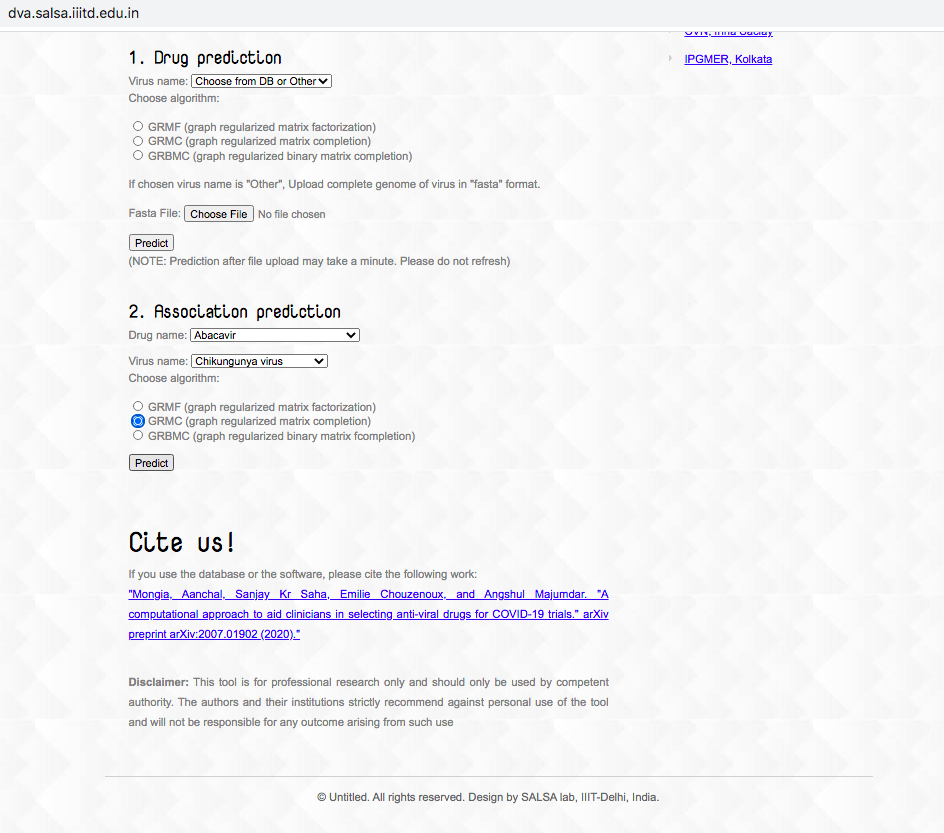}
\caption{Sample query 3- To predict association score between a drug and virus.}
\label{h6}
\end{figure}

\begin{figure}[h!]
\centering
\includegraphics[width=.98\linewidth]{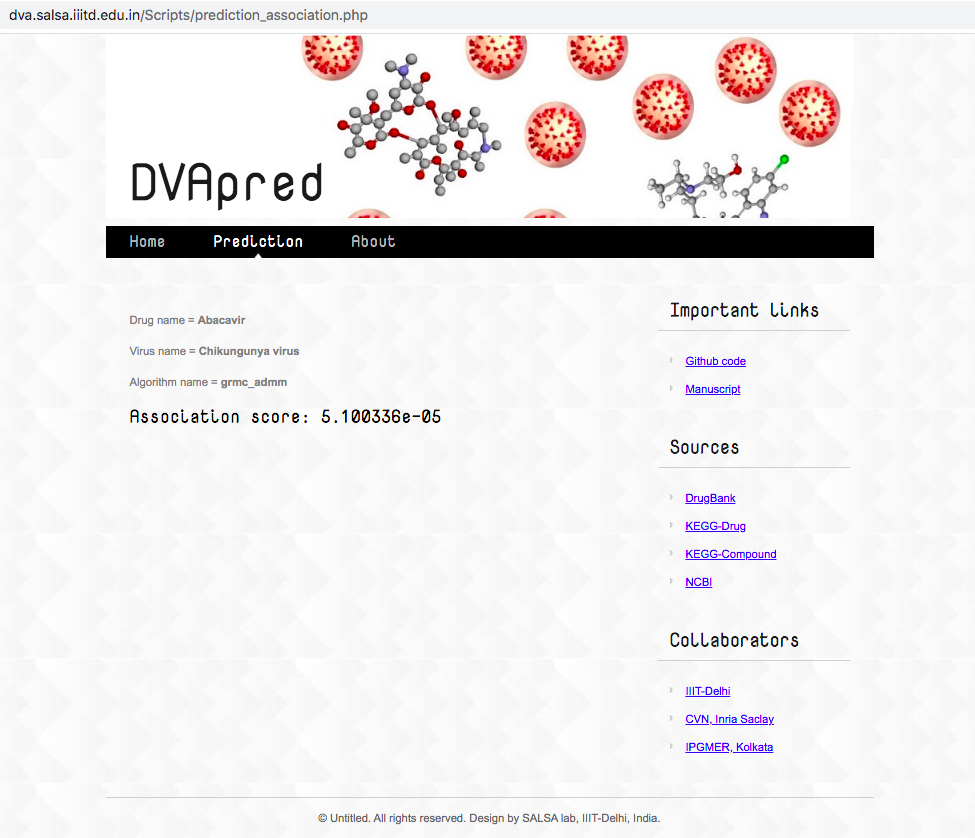}
\caption{Sample query 3- Prediction results}
\label{h7}
\end{figure}

Figure \ref{h1} shows the `Home' page of the server. Figures \ref{h2},\ref{h3},\ref{h4},\ref{h5} show how to use the web server to run an algorithm for the prediction of top-5 drugs for a virus already existing in the database (sample query 1: Figures \ref{h2} and \ref{h3}) or for a novel virus by uploading its genomic sequence in ``.fasta" file format (sample query 2: Figures \ref{h4} and \ref{h5}). Figures \ref{h6} and \ref{h7} depict the usage of the software to predict the association score (normalized) between a drug and a virus predicted by the chosen algorithm (sample query 3: Figures \ref{h6} and \ref{h7}).

\noindent \textbf{About} – This page contains of a brief description about how the dataset has been curated. It also gives short descriptions of the different algorithms used. The reference papers of the algorithms are also provided for interested readers. 

\begin{figure}[h!]
\centering
\includegraphics[width=.98\linewidth]{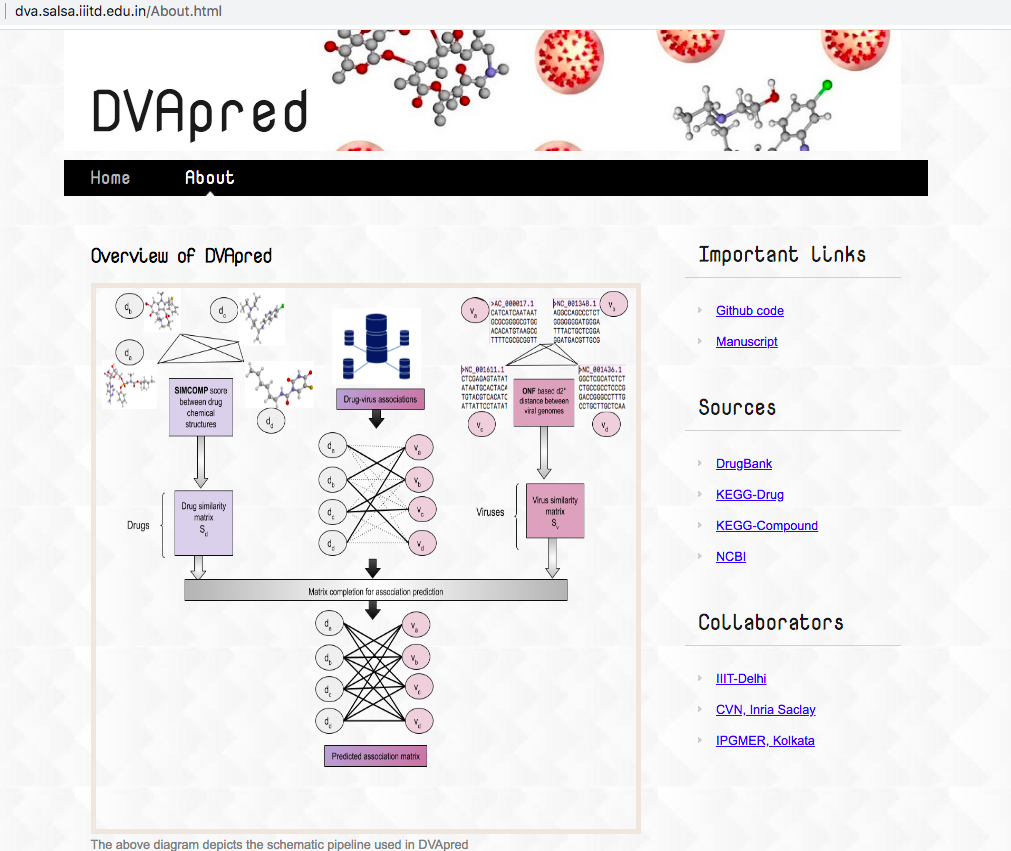}
\caption{About page of the web-server}

\label{a1}
\end{figure}


In the side menu, we also provide some important links, to our code on Github repository, our pre-print and other information pertaining to this work such as data sources and collaborations.

\end{document}